\documentclass[aps,prd,preprintnumbers,twocolumn,floats,aps,epsfig,nofootinbib,axodraw,amsmath,amssymb]{revtex4}
\usepackage{subfigure}
\def\beq{\begin{equation}}
\def\eeq{\end{equation}}
\def\bea{\begin{eqnarray}}
\def\eea{\end{eqnarray}}

\def\etmiss{E\!\!\!\!\slash_{T}}

\def\pslash{\not{\hbox{\kern-4pt $p$}}}
\def\qslash{\not{\hbox{\kern-4pt $q$}}}
\def\lv{\not{\hbox{\kern-4pt $L$}}}
\def\lsim{\mathrel{\raise.3ex\hbox{$<$\kern-.75em\lower1ex\hbox{$\sim$}}}}
\def\gsim{\mathrel{\raise.3ex\hbox{$>$\kern-.75em\lower1ex\hbox{$\sim$}}}}
\def\ifmath#1{\relax\ifmmode #1\else $#1$\fi}

\usepackage{graphicx}
\usepackage{bm}
\begin{document}

\renewcommand{\thefootnote}{\arabic{footnote}}

\preprint{UCB-PTH-09/24}

\title{Dark Matter Stabilization Symmetries and Long-Lived Particles\\
at the Large Hadron Collider}
\bigskip
\author{Devin G. E. Walker\footnote{Email Address:  dgwalker@berkeley.edu.}}
\address{Department of Physics, University of California, Berkeley, CA 94720, U.S.A,\\
Theoretical Physics Group, Lawrence Berkeley National Laboratory, Berkeley, CA 94720, U.S.A.}

\begin{abstract}
Many popular models of new physics beyond the Standard Model use a parity to stabilize weakly interacting, dark matter candidates.  We examine the potential for the CERN Large Hadron Collider to distinguish models with parity stabilized dark matter from models in which the dark matter is stabilized by other symmetries.  In this letter, we focus on signatures involving long-lived particles and large amounts of missing transverse energy.  To illustrate these signatures, we consider three models from the literature which are representative of a more general class of models with non-traditional stabilization symmetries.  The most optimistic scenario can observe the proposed signature with a minimum of 10 fb$^{-1}$ of integrated luminosity at design center of mass energy.  It will probably take considerable longer to validate the stabilizing symmetry is not a simple parity.  In all, we emphasize that the underlying symmetry that stabilizes weakly interacting dark matter has tremendous implications for the LHC and our understanding of the nature of dark matter.
\end{abstract}

\maketitle
There is an overwhelming amount of evidence for the existence of dark matter (DM).  Numerous astrophysical, cosmological and direct detection experiments~\cite{Bertone:2004pz} provide a consensus picture:  viable dark matter candidates must be stable, neutral under the Standard Model (SM), non-relativistic at redshifts of $z \sim 3000$ and generate the measured relic abundance of  $h^2\, \Omega_{\mathrm{DM}} = 0.1131 \pm 0.0034$~\cite{Hinshaw:2008kr}.  These requirements are general and upon implementation provide a plethora of models with viable DM candidates.  The most popular, e.g. supersymmetry (SUSY), Little Higgs and extra-dimensional scenarios, typically stabilize the dark matter with a generic parity ($\mathcal{Z}_2$) symmetry~\cite{Dimopoulos:1981zb,ArkaniHamed:2002qy,Cheng:2003ju,ArkaniHamed:1998rs,Randall:1999ee,Servant:2002aq,Agashe:2007jb}.  Since particles are defined by how they transform under different symmetries, in essence only one type of DM candidate is being considered!  In this letter, we consider the observational consequences of dark matter stabilized with another symmetry (a ``non-parity" symmetry) at the CERN Large Hadron Collider (LHC).  We focus on a signal of two long-lived particles plus large amounts of missing transverse energy ($\etmiss$).\footnote{In addition to long-lived particles, one can also study decay chains of particles charged under ``non-parity" stabilization symmetries.  The result is specific kinematical distributions that are distinct from traditional parity stabilized models.  We explore this research direction in the collaboration of~\cite{agashe:2009ph}.  In a separate collaboration~\cite{georgi:2009ph2}, we focus on showing how the LHC can distinguish strongly coupled hidden sectors with conformal or confined phases from perturbative ones using jets of missing energy.  Such sectors can be the source of the dark matter candidates.}  To demonstrate the viability of this signature, we devote the next section to show how such a signal is often suppressed for models with $\mathcal{Z}_2$ stabilized DM.  We next briefly review three example models in which the DM is stabilized by a symmetry other than a $\mathcal{Z}_2$; there example processes that generate the signal are provided.  Section III contains the specifics of our methodology including the SM backgrounds, detector and acceptance cuts.  Also included are the details of a fast detector simulation written to estimate the amount of missing energy generated from long-lived particles interacting with the ATLAS and CMS detectors.  The relative suppression of $\mathcal{Z}_2$ stabilized models for the signal of two observed long-lived particles plus large amounts of missing transverse energy is a central point of this paper.  This statement is quantified for an example effective theory in Section IV.  The analysis therein includes the computing of the statistical significance for the ``non-parity" signal processes from Section II.  Also considered are the well-known SM backgrounds.  Conclusions follow.

\section{Signal Suppression for Parity Stabilized Models}

In this section we provide inductive reasoning to support the assertion that models with $\mathcal{Z}_2$ stabilization symmetries are either generally suppressed or distinguishable from models with non-traditional stabilization symmetries.  These arguments are valid for signatures of two long-lived particles plus $\etmiss$.  To begin, consider the low energy effective theory of a model with parity stabilized dark matter, $\chi$.  Below the cutoff $\Lambda$, we include the SM, $\chi$ and the next-to-lightest parity odd particle, $\psi$.  We require $\psi$ to be charged under both the SM and $\mathcal{Z}_2$ symmetries.  We also require $\psi$ to have a lifetime long enough to ensure it will not decay inside of the detector yet is consistent with the strong bounds on charged, long-lived particles~\cite{Perl:2001xi}.  Now just by parity conservation, any process involving the production of parity odd particles must have an even number of such particles in the final state.  In our scenario, this means a final state involving $\psi^* \psi$, $\psi^* \psi\, \psi^* \psi$, $\psi^* \psi\, \chi^* \chi$, etc.  Now suppose a dedicated experimental analysis observes long-lived $\psi$ particles.  From our effective description, we posit final state signatures with two observed $\psi$ particles plus $\etmiss$ are generally suppressed.  Such a signature may be possible for models in which the dark matter is stabilized by other symmetries such as $\mathcal{Z}_2 \times \mathcal{Z}_2$, $\mathcal{Z}_3$, $SU(3)_\mathrm{global}$, etc.  In Section III and IV, we will account for the relevant SM backgrounds.  

To see the signal suppression in the $\mathcal{Z}_2$ case, note the most relevant operators involving couplings of $\psi$ with the SM are
\begin{align}
\mathcal{O}_{1f} = g\,\psi^c \sigma A\, \psi  && \mathcal{O}_{1b} = g\,\partial \psi^* A \psi
\label{eq:operator1} 
\end{align}
and
\begin{align}
\mathcal{O}_{2b} = \lambda_1\, h^\dagger h\, \psi^* \psi && \mathcal{O}_{2f} = {\lambda_2 \over \Lambda}\, h^\dagger h\, \psi^* \psi
\label{eq:operator2}
\end{align}
where $A$ is the gluon, photon or Z boson.  $h$ is the higgs.  Note, $h$ is general and can represent extended higgs sectors such as two higgs doublets that are common in SUSY models.  $\mathcal{O}_{if}$ ($\mathcal{O}_{ib}$) are the operators involving fermionic (bosonic) $\psi$.  The same type of couplings are also possible for $\chi$ with $A$ the photon or Z boson.  If both $\chi$ and $\psi$ are scalars, then the marginal operator
\begin{equation}
\mathcal{O}_{3b} = \lambda_3 \,\psi^* \psi \, \chi^* \chi
\label{eq:operator3}
\end{equation}
is also possible.  As well, if $\psi$ has the SM quantum numbers of the higgs then  
\begin{align}
\mathcal{O}_{4b} = {\lambda_{4b} \over \Lambda}\,\psi^* \chi h\,h^\dagger h && \mathcal{O}_{4f} = \lambda_{4f}\,\psi^* \chi\,h
\label{eq:operator4}
\end{align} 
is viable.  If $\psi$ has SM charges which allow yukawa couplings to quarks and leptons,  our assumption of a long-lived $\psi$ particle requires, $\lambda_{5,6}$, to be small for
\begin{align}
\mathcal{O}_{5} = \lambda_5 \,\psi^c \chi\, l && \mathcal{O}_{6} = \lambda_6 \,\psi^c \chi\, q.
\label{eq:operator5}
\end{align}
There may be additional higher-dimensional operators that may permit decay of $\psi$ into the SM.  We assume $\Lambda$ is sufficiently large to suppress the effects of these operators.

One way to generate the signal with our effective description is through pair production of the $\psi$ particles.  One of the $\psi$ particles can subsequently emit two $\chi$ particles via equation~\ref{eq:operator3}.  It is also possible for $\psi$ to emit a SM higgs boson or an off-shell $Z$ or photon via equations~\ref{eq:operator1} and~\ref{eq:operator2}.  These would then need to decay into two $\chi$ particles.  The first (second) scenario is suppressed by the amount $\psi$ (as well as $Z^*$/$\gamma^*$) goes off-shell.  The diagrams involving a virtual SM higgs, $h_0$, have additional off-shell suppression if $m_{h_0} < 2 m_\chi$.  To get a numerical estimate of the effect of off-shell suppression, note the integral over a Breit-Wigner resonance for one virtual particle, V, is 
\begin{equation}
I = {\theta_\mathrm{max} - \theta_\mathrm{min} \over \Gamma_V M_V}
\end{equation}
where $\theta_{\mathrm{max(min)}} =  \tan^{-1}((m_{\mathrm{max}(\mathrm{min})}^2 -  M_V^2)/\Gamma_V M_V)$.  Here $m_\mathrm{max}$ and $m_\mathrm{min}$ are the appropriate limits of the phase space integration~\cite{Han:2005mu}.  In the limit where the virtual particle is on shell $\theta_\mathrm{max} - \theta_\mathrm{min} \to \pi$.  As an example, for the process involving equation~\ref{eq:operator3}, the integral goes as $\theta_\mathrm{max} - \theta_\mathrm{min} \sim \mathcal{O}(0.01)$ for a 500 GeV $\psi$ and 100 GeV $\chi$.  The SM higgses in the diagrams described above can have heavy masses, $1$ TeV $ > m_{h_0} \geq 2\,m_\chi$, which may slightly reduce the amount of off-shell suppression; in this case often $m_{h_0} > 2\, m_Z$ and there is an additional suppression from the SM higgses preferentially decaying into W and Z bosons.  Note the partial width for $h_0 \to \chi \chi$ is
\begin{eqnarray}
\Gamma_{h_0 \to \chi \chi\,\,\mathrm{scalar}} &=& {\kappa^2_1\, v_{ew}^2 \over 16 \pi m_{h_0}}\sqrt{1 - 4 m_\chi^2/m_{h_0}^2}  \\
\Gamma_{h_0 \to \chi \chi\,\,\mathrm{fermion}} &=& { A\,\kappa^2_2 \,m_{h_0} \over 32 \pi} \biggl(1 - 4 m_\chi^2/m_{h_0}^2 \biggr)^{3/2}
\label{eq:iwidths}
\end{eqnarray}
where $\kappa_1$ ($\kappa_2$) is the coupling for scalar (fermionic) $\chi$.  Here $A = 2$, $4$ for Weyl and Dirac fermions, respectively.  The partial width for decays into W and Z bosons goes as $\Gamma \sim G_F \,m_{h_0}^3$ where $G_F$ is Fermi's constant.  Thus, as $m_{h_0}$ increases, the higgs decays more frequently into  W and Z pairs due to the $m_{h_0}^3$ enhancement.  For example, a 1 TeV (300 GeV) higgs decays into 100 GeV scalar (Dirac fermion) DM $6 \times 10^{-2}$~\% (11\%) of the time.  Here we have assumed $\kappa_i = 0.5$.  Even when the SM higgs decays at the 10\% level, an experimentalist can search for  $\psi^* \psi + h_0$ where $h_0$ decays to the SM.  As we will discuss in Section III, tagging long-lived particles is relatively easy.  Thus, precise measurement of the SM higgs branching fractions can uncover an invisible decay width.  An experimentalist can potentially identify if a $\etmiss$ signature comes from SM higgs decay.

Another way to generate the $\psi^* \psi + \etmiss$ signature is through pair producing charged higgses.  Such higgses are common in SUSY models.  The higgses decay on-shell via the operators in equation~\ref{eq:operator4} for $m_h > m_\chi + m_\psi$.  Now, in two higgs doublet models, the partial width for each higgs decay channel into the SM goes as $\Gamma \sim G_F\,m_h^3$.  Thus, from the arguments in the previous paragraph, the $h \to \psi\, \chi^*$ decay may be suppressed.  Even with a significant partial width, the experimentalist can still largely determine if higgs decay is responsible for the $\psi^* \psi + \etmiss$ signature.  The higgs decay width into $\psi\,\chi^*$  can be measured by comparing  $p\,p \to h^*\,h \to \psi + \mathrm{SM} + \etmiss $ events to the signal.  In particular, this process allows the reconstruction of both of the virtual higgses.  The critcal point:  If an experiment observes the $\psi +  \mathrm{SM} + \etmiss$ signature and the SM daughters can be reconstructed to the higgs, it is observationally clear that higgs is parity even.  If the mass of the other virtual higgs can be reconstructed, it is clear we have the case where parity even states can produce the $\psi^* \psi +\!\!~\etmiss$ signature.  The higgs mass from the $h \to \psi\, \chi^*$ decay can be reconstructed with a transverse invariant mass.  This is in analogy to $W \to l \,\nu$ mass reconstruction in $WW$ decays~\cite{Barger:1983wf}.  As before, observing the higgs decay into the SM daughters is not as problematic because the tagged  $\psi$ particles greatly reduce the SM background.  The transverse mass is defined as
\begin{equation}
M_T = \sqrt{(E_T + \etmiss)^2 - (\vec{p}_{T} + \vec{p}_T\!\!\!\!\!\!\slash\,\,\,\,)^2}
\label{eq:transversemass}
\end{equation}
where $E_T^2 = \vec{p}_T^2 + M_\psi^2$ and $\vec{p}_T$ is the transverse momentum of the $\psi$ particle.  In addition, we take $\vec{p}_T\!\!\!\!\!\!\slash\,\,\,\, = \sum \vec{p}_{T\,\,\mathrm{visible}}$ and $\etmiss^2 = \vec{p}_T\!\!\!\!\!\!\slash\,\,\,\,^2$.  As an example, we simulate the process $p\,p \to h^*\,h \to \psi \,\chi^* + \mathrm{SM}$ for a 500 GeV charged higgs and plot the reconstructed transverse invariant mass in Figure 1.  Note the transverse mass has a kinematic edge at the given higgs' mass.
\begin{figure}[t]
{\includegraphics[width=7.2truecm,height=5.7truecm,clip=true]{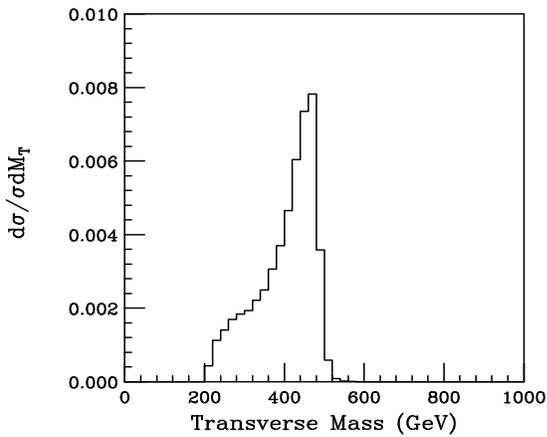}
\label{fig:transversemass}}
\caption{An example reconstructed transverse mass for $p\,p \to h^*\,h \to \psi \,\chi^* + q\, \bar{q}$ with 500 GeV charged higgs decaying into a 200 GeV long-lived $\psi$ and 100 GeV DM $\chi^*$.  The most important aspect of this plot is the kinematic edge at 500 GeV.  We apply the cuts consistent with ATLAS and CMS collaborations with a transverse momentum cut of $200$ GeV and the cuts of equations~\ref{eq:velocitycut} and \ref{eq:muoncut}.}
\end{figure}

The careful reader may argue that adding additional particles to the effective theory will generate large missing energy signatures for models with $\mathcal{Z}_2$ stabilized dark matter.  There are three types of new particles which can produce non-trivial consequences:  a particle, $\phi$, charged solely under the SM, the next heaviest parity odd particle, $\psi'$, which is also charged under the SM and $\phi_s$ a singlet under the SM and the $\mathcal{Z}_2$ symmetry.  In the first case, the coupling $\lambda_7 \,\phi^* \,\psi \chi$ is allowed; thus, generating the $\psi^* \psi +\!\!~\etmiss$ signature with the minimal amount of suppression mirrors the discussion for  the production and decay of charged higgses via equation~\ref{eq:operator4}.  We note $\phi$ cannot have random SM quantum numbers which could forbid yukawa couplings with SM particles.  This is generally true because (1) $\phi$ has the quantum numbers of $\psi^c \chi^*$ and (2) $\psi$ must decay through relevant or irrelevant operators to $\chi$ and the SM to avoid bounds on stable charged particles~\cite{Perl:2001xi}.  Like the discussion of the two higgs doublets,  $\phi$ should decay more prominently to the lighter SM particles than $\psi$ and $\chi$; therefore, the  $\psi^* \psi + \etmiss$ signal may be suppressed.  As before, if there is a non-trivial decay width into $\psi\,\chi^*$, plotting the transverse invariant mass could provide identification of $\psi^* \psi + \etmiss$ production from a parity even $\phi$.  Consider now two new effective descriptions where we separately add $\psi'$ and $\phi_s$ to the original effective theory of the SM, $\psi$ and $\chi$.  In the $\psi'$ effective theory, it is clear that there are no $\psi' \psi\, \chi$ couplings;  however, if $\psi$ and $\psi'$ have the same charges under the SM, they can mix by an angle, $\theta$, that is a function  of the mass ratio $m_\psi/m_{\psi'}$.  Any cross section involving pair production of a heavy $\psi'$ with a $\psi^*\, \psi\, \chi^* \,\chi$ final state is suppressed by at least $\theta^4$.  We have not demanded that $\psi'$ is long lived and it can conceivably decay to the SM and $\chi$ via operators analogous to equation~\ref{eq:operator4}; thus, a small branching fraction into our signal may be possible.  In the  $\phi_s$ effective theory, $\phi_s$ must have $\mathcal{O}(1)$ couplings to the SM in order to be produced at the LHC; thus, it generally mirrors the SM higgs and charged higgs discussions above.  The signal could be generated from $\phi_s$ pair production and decay via $\phi_s\,\psi^* \psi$ and $\phi_s\, \chi^* \chi$ couplings.  Like before, the parity charge of $\phi_s$ can be ascertained by considering the  $\phi_s\phi_s \to \chi^*\,\chi + \mathrm{SM}$ process.  Both $\phi_s$ masses could be reconstructed in analogy to  $Z Z$ decay to two neutrinos and two visible SM particles~\cite{Barger:1987du}.  The reader might now argue that a new symmetry or large coupling could be implemented so that $\phi_s$ decays solely to $\psi$ and DM pairs.  Likewise, a similar argument could be constructed so $\phi$ decays solely to DM and $\psi$.  The large coupling case implies a new accidental symmetry.  In both cases, $\psi$ and $\chi$ would then have to be charged under these new symmetries.  This is tantamount to the new stabilization symmetry we sought to distinguish in the first place.  

An effective theory with some number of $\phi$, $\psi'$ and $\phi_s$ particles is also possible.  Given all of the arguments above, we assert such an effective theory either generates a suppressed signal or can be identified as a scenario with a $\mathcal{Z}_2$ stabilizing symmetry.  If further convincing is necessary, one can list all of the tree-level decay topologies which result in a $\psi^*\, \psi\, \chi^* \,\chi$ final state where the mother particles have non-specific charges.  The least suppressed diagrams from this exercise are analogous to examples above.  As a final note, since both $\phi$ and $\phi_s$ can potentially have tree-level SM couplings, generally precision electroweak measurements place constraints on their mass to roughly 5 TeV~\cite{bounds}.  In the next section, we briefly outline three models in the literature in which the dark matter is stabilized by a non-traditional ``non-parity" symmetry.  We will subsequently show how these models generate relatively large $\psi^* \psi + \etmiss$ signatures.  Ultimately, our signal will be a statistically significant  excess of missing energy in this channel.  We present each model's results in comparison to an effective theory with a $\mathcal{Z}_2$ stabilization symmetry.  This comparison effective theory has a particle content of the SM, the SM higgs, $\psi$, $\psi'$ and Dirac fermion DM.  The missing energy is generated from the process $h_0 \to \chi^* \chi$ for a SM higgs mass of $350$ GeV.  The DM is considered to be Dirac DM in order to maximally enhance the invisible branching fraction.

\section{Example Models}

In this section, we briefly outline the relevant aspects of three models in the literature that can generate our signature.  The first model stabilizes the dark matter with a linearly realized discrete symmetry generated from spontaneous symmetry breaking.  The second stabilizes with a $\mathcal{Z}_3$ symmetry generated from baryon number.  The final model has DM stabilized by a parity symmetry.  A light, dark gauge boson which couples to the DM helps to mediate DM annihilation.

\subsection{A Companion Model}

In a companion paper, we present a class of models in which spontaneous symmetry breaking is used as a mechanism to naturally correlate the weak scale and the mass scale associated with the dark matter annihilation cross section. See~\cite{walker:2009md} for details.  The featured model has a $SU(2)_D$ that commutes with the SM; the symmetry is ``broken" along with electroweak symmetry through an extended higgs sector to generate a linearly realized $\mathcal{Z}_2$ symmetry.  The result is new dark gauge bosons with a mass of order the weak scale.  These models can accommodate additional messenger vector-like heavy fermions, $Q$, whose mass is, for our purposes, a relatively free parameter~\cite{Perl:2001xi}.  $Q$ is charged under the SM as well as the $SU(2)_D$.  For our signature, we consider these vector-like fermions, $Q$, as long long lived.  We are interested in the effective operators
\begin{equation}
\mathcal{O}_{7} = g' Q^* \sigma A \,Q 
\label{eq:operator10}
\end{equation}
where $A$ is the new gauge bosons that does not couple directly to the SM.  As shown in~\cite{walker:2009md}, the new gauge bosons decay into two DM candidates, $\chi$, with a 100\% branching fraction via the operator
\begin{equation}
\mathcal{O}_{8} = g' \chi^* \sigma A\,\chi 
\label{eq:operator11}
\end{equation}
Thus, the signal for this model at lowest order is $p\,p \to Q \,Q^* \, A \to Q \,Q^* \,\etmiss$.  There is a seemingly irreducible background of $p\,p \to Q \,Q^*$.  We will show how to distinguish this background in the following sections.  Further, we also show a large excess of missing energy in this channel in comparison to the $\mathcal{Z}_2$ scenario described in the previous section.  In the coming analysis, we refer to $Q$ and $A$ as $\psi$ and $\chi$, respectively, in order to be consistent with the notation in previous sections.

\subsection{Agashe and Servant}

Agashe and Servant (AS) proposed a class of grand unification models within the framework of warped extra dimensional space-time.    Please see~\cite{AS} for more details.  In these models, the SM along with new exotic matter fills out the $SO(10)$ spinor representations.  Each spinor multiplet is charged under baryon number.  Notably, all of the non-SM particles transform under a $\mathcal{Z}_3$ symmetry generated from the particles' color and fractional baryon number
\begin{equation}
\eta \to \exp{\biggl[2\pi i \biggl(B - {n_c - \bar{n}_c \over 3}\biggr)\biggr]} \eta.
\end{equation}
Here $\eta$ is a generic of any particle in the theory and $B$, $n_c$ ($\bar{n}_c$) are the particle's baryon number and number of colors (anti-colors).  A natural DM candidate, $\nu_R$, with the SM charges of a right-handed neutrino, is in same the spinor multiplet as the SM right-handed top, $t_R$, with $B = -1/3$.  Also of interest to us is an exotic, $SU(2)_L$ doublet fermion, $Q_L$, which transforms as the fundamental under color $SU(3)$.  $Q_L$ is charged under the $\mathcal{Z}_3$ and is in the same spinor multiplet as $\nu_R$; thus, it can be relatively light compared to the other new exotic states.  Because the $SO(10)$ is broken to the SM gauge group, there are additional gauge bosons beyond the SM that are charged under the SM and $Z_3$ symmetries.  The $X$ boson has the same SM quantum numbers as $Q_L$ but with a different baryon number of $B = -2/3$.  We note because the SO(10) was broken with varying boundary conditions than the SM, proton decay constraints are eliminated.  We focus on the coupling
\begin{equation}
\mathcal{O}_{9} = g_5\, Q_L^*\, \sigma \,\nu_R \, X
\label{eq:operator8}
\end{equation}
where $g_5$ is the Pati-Salam gauge coupling.  There is an additional Pati-Salam gauge boson, $X_s$, which is important when considering the decay of $Q_L$ to the SM and $\nu_R$.  This $Q_L$ decay can only occur through equation~\ref{eq:operator8} where the $X$ boson mixes with the $X_s$ boson.  The $X_s$ boson subsequently decays to $\nu_R$ and $t_R$ via
\begin{equation}
\mathcal{O}_{10} = g_5\, t_R^* \,\sigma\, \nu_R \, X_s
\label{eq:operator9}
\end{equation}
The mixing between the $X$ and $X_s$ bosons occurs through a vaccum expectation value (vev) that can be effectively dialed.  We assume a value that generates long-lived particles consistent with~\cite{Perl:2001xi}.  Consider the AS model where the SO(10) gauge group is broken on the infrared brane.   In that scenario,  the X bosons can be relatively light.\footnote{We thank K. Agashe for alerting us to this point.}  In the case where the SO(10) is broken on the UV brane, the $X$ bosons must be heavier than about 3 TeV.  We consider the former and take the process $p\,p \to Q_L\, Q_L^* \to  \nu_R\, \nu_R^*\, X\, X^*$ with $X$ being long lived as our signal.  We consider only QCD production involving
\begin{equation}
\mathcal{O}_{11} = g_s\,Q_L^*\, \sigma G \,Q_L  
\end{equation}  
where $G$ is the gluon.  We take the mass hierarchy of $m_{Q_L} > m_X + m _{\nu_R}$.  The signal is enhanced because $Q_L$ decays to $X^*\,\nu_R$ with a 100\% branching fraction.  In the analysis section, we refer to $X$ and $\nu_R$ as $\psi$ and $\chi$, respectively, in order to be consistent with the notation in previous sections.

\subsection{Light Hidden Dark Sectors}

Pospelov, Ritz and Voloshin proposed a class of models in which the DM has ``secluded" interactions with the SM.  See~\cite{Pospelov:2007mp} for details.  This is done so their models can be consistent with increasingly stringent bounds from elastic DM scattering with nuclei in direct (as well as some indirect) detection experiments.  The DM annihilates through metastable mediators which in turn eventually decay to the SM.  All of the secluded versions of their models have the characteristic that the DM mass is larger than the mediator mass.  As an example, they construct a model in which a new U(1) gauge boson kinetically mixes with SM hypercharge.  After spontaneous symmetry breaking, the new gauge boson gets a mass and has a lifetime which, for most of the parameter space in the model, is stable on detector time scales.~\footnote{We thank A. Nelson for this observation.}  This model differs from the model in Subsection IIA in that the former uses spontaneous symmetry breaking to generate the dark matter stabilization symmetry.

More recently, Arkani-Hamed, Finkbeiner, Slatyer and Weiner (AFSW) proposed a class of models to provide a dark matter interpretation for the PAMELA anomaly. The PAMELA collaboration searches for indirect signals of dark matter by observing high-energy cosmic rays from orbit.  The collaboration observes an anomalously large number of positrons in the cosmic ray fluxes which do not conform to the theoretically expected background.  In constrast, the anti-proton to proton flux ratio does conform to theoretically expected background~\cite{Pamela}.  The authors proposed a class of models where a new force helps to mediate the DM annihilation.  See~\cite{AFSW} for details.  To be consistent with the positron and anti-proton measurements, the new gauge boson(s) must have a mass less than twice the mass of the proton.  Thus, kinematically, the new gauge bosons therefore can only decay into the lighter leptons.  The fact that this new gauge boson couples to leptons illustrates a primary difference between this and the companion model in the first subsection.  

For both scenarios, we assume a new light gauge boson with a mass of 1 GeV.  The gauge boson does not decay  inside the detector and constitutes the missing energy.  This is in analogy to the signal process in Subsection IIA.  For the AFSW model, this means we assume the new gauge boson to be long lived and not decay into leptons inside the detector.\footnote{We thank N. Weiner for suggesting this point for the model of Arkani-Hamed, et al.}  We also assume a vector-like quark production mechanism via the effective operator in equation~\ref{eq:operator10}.

\section{Methodology}

A huge key to our signature are the particles, $\psi$, which are long lived and charged under both the SM and dark matter stabilization symmetry(-ies).  Here we briefly review some important aspects of the collider physics of long-lived particles.  A review can be found in~\cite{sreview}.  

If $\psi$ is charged under $SU(3)$ color, it, after production, forms color singlet states with valence SM quarks.  These $\psi$-hadrons interact with the detector by exchanging SM quark partons with the material therein.  This nuclear scattering process causes these $\psi$-hadrons to potentially charge flip from charged to neutral (and vice-versa), lose energy and/or even stop inside the detector.  The heavy $\psi$ parton inside the $\psi$-hadron carries a dominant fraction of the total energy of the system with the interacting SM quark partons carrying approximately $\mathcal{O}(0.001)$ fraction of the rest.  Thus, the nuclear interactions of these $\psi$-hadrons are dominantly low-energy QCD with the changes in momentum due to interactions with the detector small relative to the total momentum.  If $\psi$ is not colored, then it interacts only electromagnetically with the detector.  In this and in the charged $\psi$-hadron case, a highly ionized track is produced in the detector's tracking chamber.  Upon reaching the muon detector the charged particles look like a slow, heavy ``muon."

We are interested in large $\etmiss$ signatures generated by the dark matter candidates.  Since the hadrons can potentially charge flip to neutral as well as lose energy through interactions with the detector, it is important to quantify the amount of missing energy in each event.  To address this, we wrote a fast detector simulation which parametrizes the charge flipping and energy loss effects accounted for in GEANT3 and 4~\cite{geant}.  In our simple setup, we assume perfect lead calorimeters.  The hadron interactions therein are assumed to be perfectly opaque while the signals of the interactions are perfectly translucent.  The detector response is extrapolated from SUSY gluino R-hadron nuclear interactions for iron~\cite{Kraanetc}.  We do not include ionization effects in our simulation; such dominates over nuclear interactions only when the hadrons are at very small velocities.  The geometry for the ATLAS and CMS detectors in terms of nuclear interaction lengths is taken from~\cite{jinst}.  
\begin{figure}[t]
{\label{fig:misset1}
	\includegraphics[width=7truecm,height=5.2truecm,clip=true]{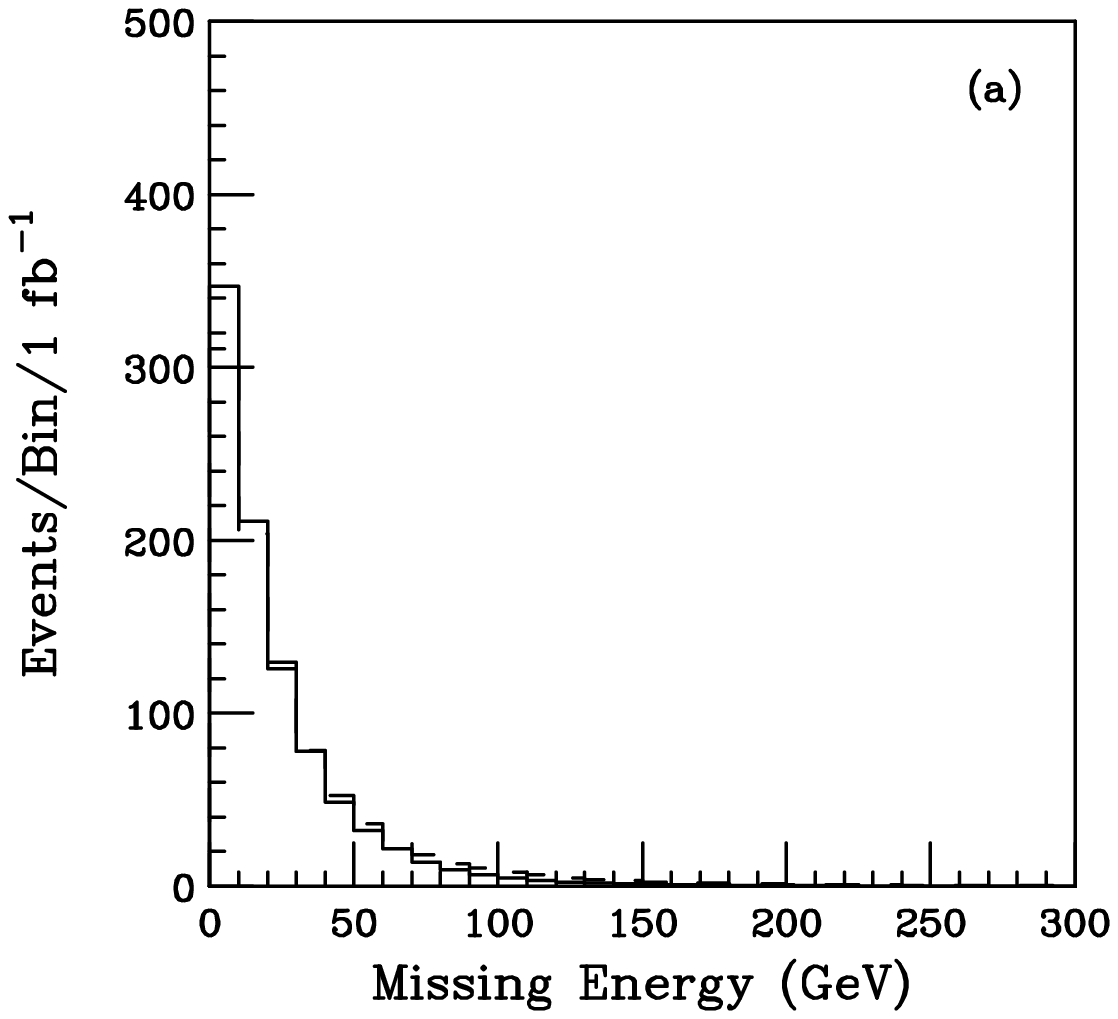}}
\vspace{0.3cm}
{\label{fig:misset2}
	\includegraphics[width=7truecm,height=5.2truecm,clip=true]{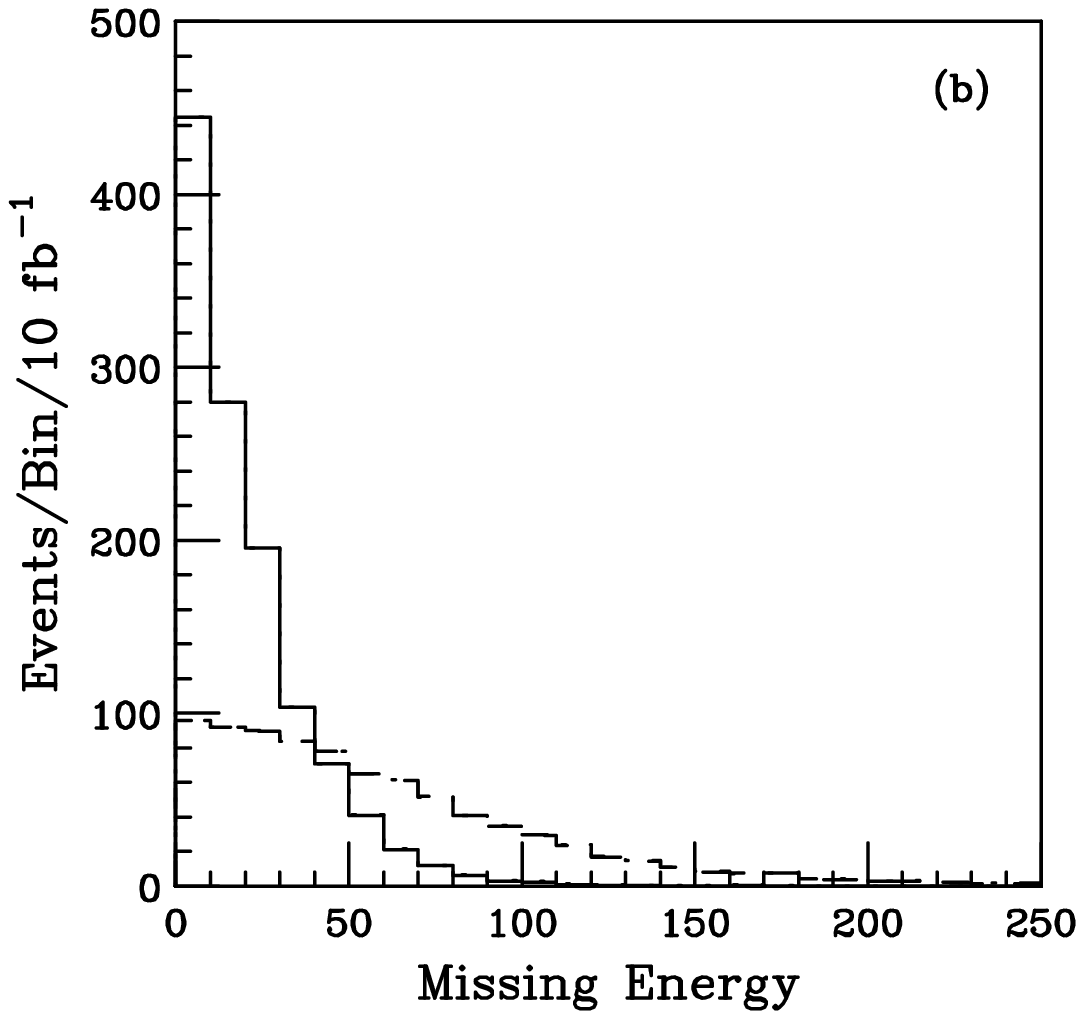}}
\vspace{0.3cm}
{\label{fig:etmiss}	
	\includegraphics[width=6.6truecm,height=5.2truecm,clip=true]{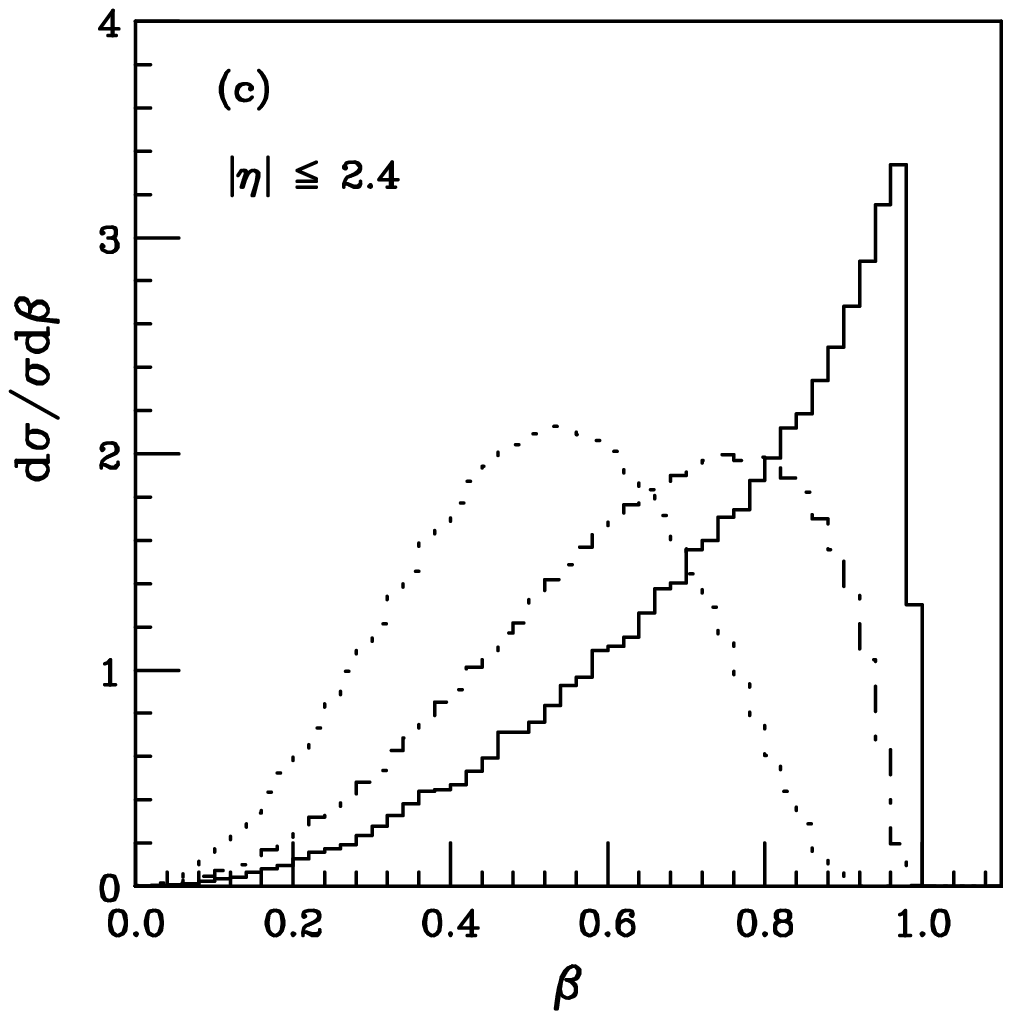}}
\caption{(a) Events vs. missing energy ($\etmiss$) for two 300 GeV quarks after transversing the simulated ATLAS (dashed) and CMS (solid) inner detector and calorimeters.  (b) The same plot as (a) but with the kinematic cuts from equations~\ref{eq:velocitycut}-\ref{eq:muoncut}.  Also shown for ATLAS (dot-dashed) and CMS (dotted) is the $\psi^* \psi + \etmiss$ signal in addition to the $\psi^* \psi$ background for the companion model in section IIA.  $g'$ is set to be electroweak strength. (c)  The normalized cross section versus velocity for a 300 (solid), 900 (dashed) and 2000 (dot-dashed) GeV long-lived vector-like quarks in the central detector region.}
\end{figure}
In Figure 2a, we plot the missing energy generated from two tagged $\psi$ particles after interacting with the calorimeters in our ATLAS and CMS detector simulation.  The $\psi$ particles have mass of 300 GeV and transform as a fundamental under color SU(3).   The similarity between the two curves is due to the similarity of the ATLAS and CMS calorimeters and inner detectors in nuclear interaction lengths.  It is comforting that the resulting distribution favorably approximates the ATLAS GEANT3 simulation for stable gluinos in Figure 9 of~\cite{Kraan:2005ji}.  In Figure 2b, we plot the missing energy for the $p\,p \to \psi^* \psi + \etmiss$ signal as well as the $p\,p \to \psi^* \psi$ background for 10 fb$^{-1}$ of luminosity.  We used the companion model scenario described in the previous section as well as the parameters listed in Section IV.   We also used the ATLAS and CMS cuts described in the next paragraph.  Here the difference between the ATLAS and CMS curves is due to the differing kinematic cuts.  It is clear the missing energy generated from the long-lived particles traveling through the detector has the potential to obscure the signal.  This plot demonstrates the necessity of writing the fast detector simulation. 

Any SM process which can generate muon tracks and/or hard jets that punch through to the muon detector provides a potential background to our signal.  To begin, it should be noted the (near-)massless particles have a velocity, $\beta$, near speed of light ($\beta = 1$); the heavier hadrons travel at much slower velocities and arrive at the muon detector ``out of time" with the rest of the event.  In figure 2c, we plot the velocity for hadrons of different masses.  It is clear most of the events have $\beta < 1$.  An additional complication occurs if the long-lived particle arrives too late.  Each proton beam at the LHC is collimated in bunches and is scheduled to be nominally separated by 25 nanoseconds~\cite{Evans:2008zz}.  Very slow hadrons could be mis-reconstructed with the wrong event.  Further, without the proper cut, cosmic ray muons could provide an additional background.  Because up to three events could be in the detector at one time, the larger ATLAS detector requires a higher lower velocity cut.  We adopt velocity cuts~\cite{expcuts} suggested by the ATLAS and CMS collaborations
\begin{align}
0.7 < \beta_\mathrm{ATLAS} < 0.9 && \mathrm{and} && 0.5 <  \beta_\mathrm{CMS} < 0.8.
\label{eq:velocitycut}
\end{align}
To limit the effect of hard jets punching through to the muon detector and faking the long-lived particles, the ATLAS collaboration adopted additional stringent $p_T$ cuts~\cite{expcuts} 
\begin{align}
p_{T\,\,\mathrm{ATLAS}} > 250\,\,\,\mathrm{GeV} && \mathrm{and} && p_{T\,\,\mathrm{CMS}} > 30\,\,\,\mathrm{GeV}. 
\label{eq:pTcuts}
\end{align}
For the ATLAS (CMS) cut, at least one (both) of the ``muons"  must satisfy the cut.  The smaller CMS $p_T$ cut is due to the collaboration requiring both a highly ionizing track in the inner detector as well as muon detector information to eliminate SM backgrounds.  This ionized track is subject to the velocity cut in equation~\ref{eq:velocitycut}.  Given the measured velocity and momentum, CMS also requires the average reconstructed particle's mass, $m$, to be
\begin{equation}
m_\mathrm{CMS} > 100\,\,\,\mathrm{GeV}.
\label{eq:invcut}
\end{equation}
where $P = \beta m /\sqrt{1 - \beta^2}$.  Full simulation of ionization effects are beyond the scope of this work.  As a simplification, we simply retained all events with charged hadrons in the simulated CMS tracker.  For the simulated events we tag at least two long-lived particles in the muon detector.\footnote{If an odd number of long-lived, visible particles are detected in addition to the $\psi^* \psi + \etmiss$ signature, it is clear a symmetry other than a parity is responsible.}  We require these particles to be in the triggerable part of the muon detector
\begin{align}
|\eta | \le 2.4.
\label{eq:muoncut}
\end{align}
ATLAS  requires no hard jets in the calorimeters to come within a cone of $\Delta R_\mathrm{ATLAS} \le 0.4$ of the ``muon" track.  CMS requires the track from the inner detector to be in a $\Delta R_\mathrm{CMS} \le 0.1$ cone of the ``muon" track in the muon detector.  Here $\Delta R = \sqrt{\Delta \eta^2 + \Delta \phi^2}$ where $\eta$ ($\phi$) is the pseudorapidity (transverse angle) of the event.  For the ATLAS cuts, the collaboration finds a signal to background ratio of $S/B = 2.6 \times 10^3$ for an integrated luminosity of 1 fb$^{-1}$.  Similarly, for the CMS cuts, that collaboration find a ``background free region" for 100 pb$^{-1}$~\cite{expcuts}.  With such a high expected $S/B$ ratio from both collaborations, our analysis will not focus on the traditional SM backgrounds.  

Because we are looking for signatures with at least two long-lived particles plus large $\etmiss$, there are irreducible backgrounds to consider.  To begin, the Z boson can decay to neutrinos and generate the signal via $p\,p \to \psi^* \psi \,Z \to \psi^*\, \psi \,\nu^* \,\nu$.  The precise Z invisible width is well known and can be accounted for.  Further verification the Z is the cause of the missing energy can come from counting the number of $pp \to \psi^* \psi \,Z \to \psi^* \psi \,\bar{l} \,l$ or $\to \psi^* \psi \,\bar{q}\, q$ events and reconstructing the $Z$ invariant mass for identification.  To be conservative, in addition to the invisible Z width, we add an invisible SM higgs decay width.  We assume the operator, $\kappa\,h\,\chi^* \chi$, where $\kappa \approx 0.5$.  $\chi$ is taken to be Dirac dark matter to maximize the invisible decay width.  See equation~\ref{eq:iwidths}.  We require a statistically significant number of signal events above these irreducible backgrounds.  Finally, especially for the first and third models in the previous section, pair production for $\psi$ particles dominates typically by orders of magnitude over the $\psi^* \psi + \etmiss$ signature.  The $\psi^* \,\psi$ production is generally back-to-back.  To further define the signal, we require the angle in the transverse plane to be
\begin{equation}
\cos \phi > -0.9.
\label{eq:cosphicut}
\end{equation}
to effectively eliminate the $\psi^* \psi$ production.

\section{Analysis and Results}

In this section, we show a statistically significant excess of missing energy for our signal of two long-lived particles, $\psi$, plus large amounts of missing transverse energy, $\etmiss$.  To characterize this signal, we focus on the three models described in Section II.  All events are simulated for the LHC at design 14 TeV center of mass energy. Our simulations use CTEQ  Set 4M parton distribution functions.  $\alpha_s$ is calculated at two loops with the renormalization and factorization scales set by $\sqrt{\hat{s}}/2$.  After implementing the cuts described in the previous section, we assume 100\% tagging efficiency for the visible heavy, long-lived particles.  The masses of the stable particles are taken as
\begin{equation}
m_\psi = 300, \,600\,\,\mathrm{GeV}.
\label{eq:masses}
\end{equation}
The mass of the DM candidates is set to be
\begin{equation}
m_\chi =100\,\,\mathrm{GeV}.  
\end{equation}

In the first section, we argued models with parity stabilized dark matter generated suppressed $\psi^* \psi + \etmiss$ signatures.  To quantify this statement we consider a parity conserving effective theory.  As discussed before, we choose the effective theory to have the SM, SM higgs and parity odd $\psi$ and Dirac DM, $\chi$, as the relevant low energy degrees of freedom below the cutoff $\Lambda$.  We require the SM higgs to have a significant invisible branching fraction into the DM.  Conservatively, we take the coupling from equation~\ref{eq:operator2} to be $\lambda_{2f} = 2$ with $\Lambda = 1$ TeV.  As discussed below equation~\ref{eq:iwidths}, a 300 GeV SM higgs decaying into Dirac DM has a relatively enhanced invisible width.  For these parameters, the generated background events are listed in Table I.  As shown, the process involving the invisible higgs decay width is generally down by a factor of 100 from the process involving a $Z$ invisible width.  This is primarily due to the amount the long-lived $\psi$ must go off-shell to accommodate the decay.  Further, since the SM higgs couples to all of the massive SM particles and must be heavier than twice the DM mass in order for the process to avoid additional off-shell suppression, we remind the reader this suppression is a general feature.  We now focus solely on the invisible $Z$ background.
\begin{table}[t]
{\footnotesize
\begin{tabular}{|c|c|c|c|} 
\multicolumn{4}{c}{\normalsize{Background Events}} \\  \multicolumn{4}{c}{ } \\  
\hline\hline
ATLAS				& 300 GeV $\psi$				&  600 GeV $\psi$ 			&  300 GeV $\psi$  \\ 
					&	\multicolumn{2}{c|}{(with $\cos \phi$ cut)} 					&  (without $\cos \phi$ cut)  \\ \hline 
$Z \to \nu^* \nu$		& 71.2  						&  6.1					&  130 \\
$h_0 \to \chi^* \chi$ 		& 0.126						& 0.01					&  0.183 \\ \hline
CMS					&							& 						&  \\  &&& \\ \hline
$Z \to \nu^* \nu$  		&  49							&  5.4					& 73.6 \\
$h_0 \to \chi^* \chi$  		& 0.055						& 0.006					& 0.066 \\ \hline \hline
\end{tabular}}
\caption{The simulated background for 10 fb$^{-1}$ of integrated luminosity.  Here the $Z \to \nu^* \nu$ and $h_0 \to \chi^* \chi$ labels correspond to the background processes: $p\,p \to\psi^* \psi\, Z \to \psi^* \psi \,\nu^* \nu$  and $p\,p \to \psi^* \psi\, h \to \psi^* \psi \,\chi^* \chi$.  The former is the well known invisible $Z$ decay; the latter is the assumed invisible higgs decay width for the parity odd effective theory described below.  $\chi$ is taken to be a 100 GeV Dirac DM.  The larger suppression $h_0 \to \chi^* \chi$ is due primarily to the heavier 300 GeV SM higgs forcing the $\psi$ off-shell as well as the 11\% invisible branching fraction.}
\label{table:1}
\end{table}

\subsection{A Companion Model}

\begin{figure}[t]
{\label{fig:costheta1}
	\includegraphics[width=7truecm,height=5.2truecm,clip=true]{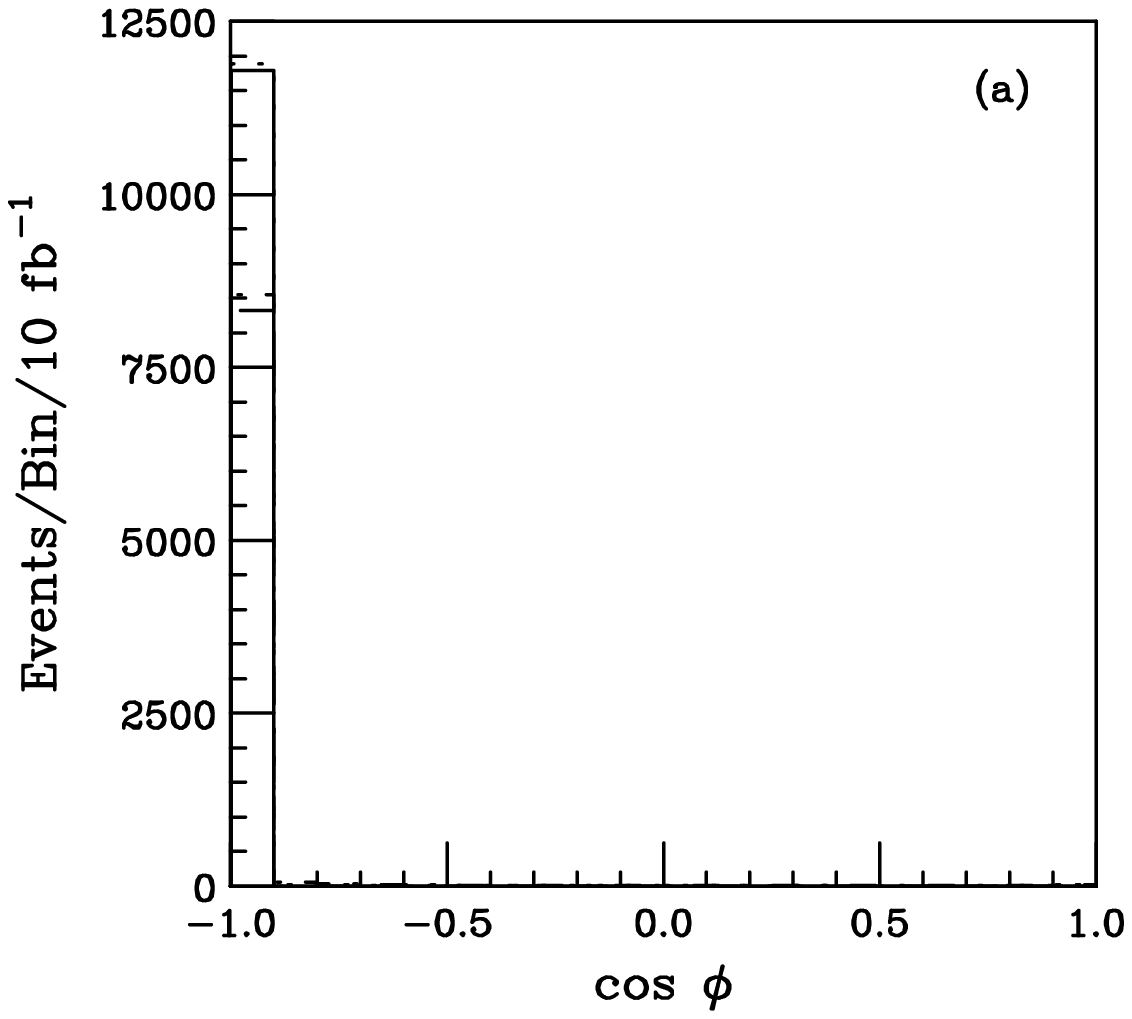}}
	\vspace{0.3cm}
{\label{fig:costheta2}
	\includegraphics[width=7truecm,height=5.2truecm,clip=true]{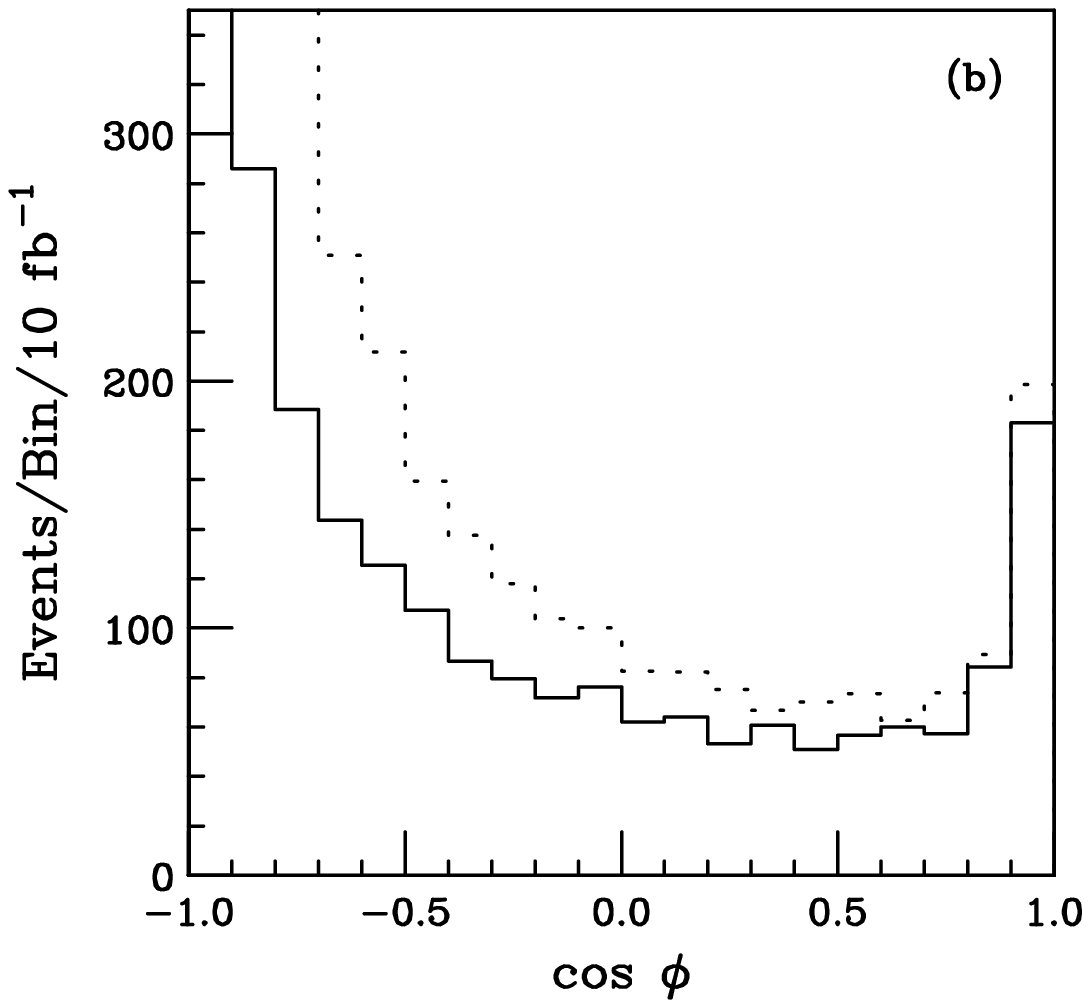}}
	\vspace{0.3cm}
{\label{fig:etmiss}	
	\includegraphics[width=7.1truecm,height=5.2truecm,clip=true]{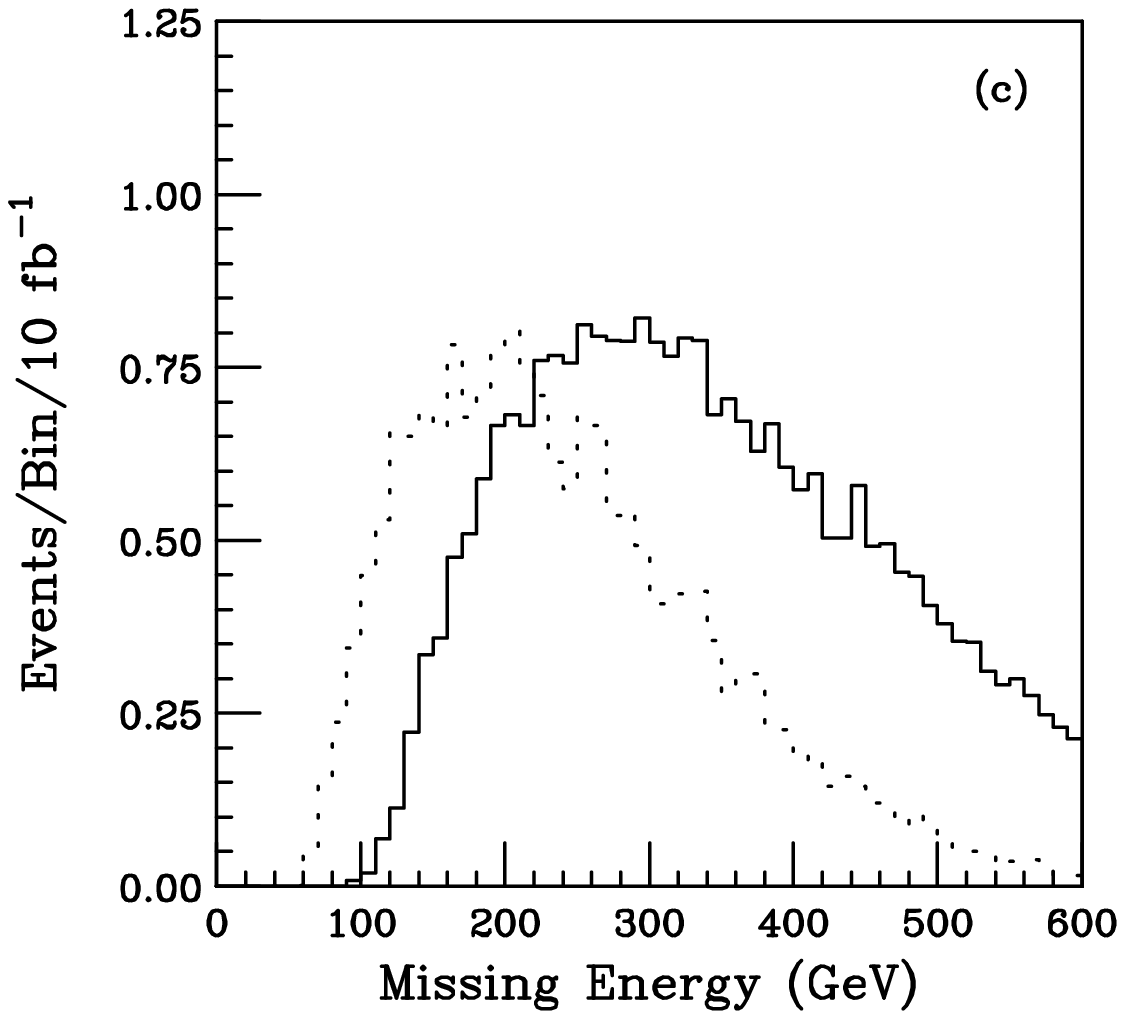}}
\caption{(a) Events vs. \!\!\!$\cos \phi$ for ATLAS (solid) and CMS (dashed) for the $\psi^* \psi$ background along with the $\psi^* \psi \,\etmiss$ signal.  The missing energy is from the emission of a $100$ GeV heavy gauge boson as described in Section IIA.  The $\psi$ mass is 300 GeV.  (b) The same plot as the first but rendered to illustrate the signal events.  (c) For the model in Section IIA, signal events vs. \!\!$\etmiss$ with the $\cos \phi < -0.9$ cut for ATLAS (solid) and CMS (dotted).  All the plots implement the kinematic cuts in Section III.}
\end{figure}
The dark gauge coupling in the companion model is taken to have the same strength as the electroweak coupling measured at the $Z$ mass $g' = g_\mathrm{ew}(M_Z)$.  In Figure 3a, we plot the cosine of the angle between the reconstructed $\psi$-hadron momentum in the plane transverse to the beam direction.  This is done for the $p\,p \to \psi^* \psi$ background as well as the $p\,p \to \psi^* \psi\,\chi$ signal for both the ATLAS and CMS detectors.  We remind the reader $\chi$ ($\psi$) is the new heavy gauge boson (vector-like quarks) described in Section IIA.  It is clear the missing energy signal is swamped and most of the events are back-to-back in the transverse plane.  Please also refer to Figure 2b; there we plotted the reconstructed missing energy from the $\psi$-hadrons.  The missing energy signature is also swamped.  In Figure 3b, we redisplay the same plot to emphasize the signal events.  The dark matter ``kicks" the reconstructed $\psi$-hadrons from being exactly back-to-back.  After applying the $\cos \phi$ cut, equation~\ref{eq:cosphicut}, none of the  $p\,p \to \psi^* \psi$ events remained.  We plot the residual signal missing energy events in Figure 3c for ATLAS (solid) and CMS (dotted).  The generated signal events for 300 GeV long-lived $\psi$ and 100 GeV DM for 10 fb$^{-1}$ of integrated luminosity is 
\begin{align}
S_\mathrm{ATLAS} = 290 	&& S_\mathrm{CMS} =  192 .
\end{align}
For a 600 GeV long-lived $\psi$ with the same luminosity, the generated signal cross section is
\begin{align}
S_\mathrm{ATLAS} = 25	&& S_\mathrm{CMS} =   22.
\end{align}
In Table II, we summarize the signal significance as well as the signal-to-background ratio for all three example models.  For the companion model, it is clear the signal can be observed for the chosen parameters.
\begin{table}[t]
{\footnotesize
\begin{tabular}{|c|c|c|c|c|c|} 
\multicolumn{5}{c}{\normalsize{Companion Model}} \\  \multicolumn{5}{c}{}  \\
 \multicolumn{1}{c}{ }		& \multicolumn{2}{c}{ATLAS}	&  \multicolumn{2}{c}{CMS}	\\ \hline \hline
			& $S/\sqrt{B}$   		& $S/B$ 	  	& $S/\sqrt{B}$ 		& $S/B$   			\\   \hline 
300 GeV $\psi$	& 34.3			& 4.1			& 29.4	& 3.92		\\  
600 GeV $\psi$	& 10.1			& 4.1			& 9.5		& 4.1	 		\\ \hline \hline  
\end{tabular}
\begin{tabular}{|c|c|c|c|c|} 
\multicolumn{5}{c}{ }\\
\multicolumn{5}{c}{ }\\
\multicolumn{5}{c}{\normalsize{Agashe-Servant}} \\  \multicolumn{5}{c}{ } \\
 \multicolumn{1}{c}{ }		& \multicolumn{2}{c}{ATLAS}	&  \multicolumn{2}{c}{CMS}	\\ \hline \hline
			& $S/\sqrt{B}$   		& $S/B$ 	  	& $S/\sqrt{B}$ 		& $S/B$   			\\   \hline  
300 GeV $\psi$	& 34.3			& 4.1			& 29.4	& 3.92		\\  \hline \hline  
\end{tabular}
\begin{tabular}{|c|c|c|c|c|} 
\multicolumn{5}{c}{ }\\
\multicolumn{5}{c}{ } \\
\multicolumn{5}{c}{\normalsize{Light Hidden Sectors}} \\  \multicolumn{5}{c}{}  \\
 \multicolumn{1}{c}{ }		& \multicolumn{2}{c}{ATLAS}	&  \multicolumn{2}{c}{CMS}	\\ \hline \hline
			& $S/\sqrt{B}$   		& $S/B$ 	  	& $S/\sqrt{B}$ 		& $S/B$   			\\   \hline 
300 GeV $\psi$	& 5.33			& 0.63		& 8.76	& 1.04		\\  
600 GeV $\psi$	& 0.75			& 0.30		& 1.08	& 0.47	 		\\ \hline \hline  
\end{tabular}
}
\caption{The signal significance and signal-to-background ratio for the described models in Section II for 10 fb$^{-1}$ of integrated luminosity.  Included is the dominant invisible Z background.}
\label{table:2}
\end{table}

\subsection{Agashe and Servant}

In this analysis, we take the Pati-Salam coupling to be $g_5 = 0.7$ with $m_{Q_L} = 600$ GeV.  The long-lived particle, $\psi$, is the heavy Pati-Salam gauge boson with a $300$ GeV mass; recall, this $X$ boson is a color triplet and picks up valence quarks to form the long-lived $\psi$-hadron.  Because $Q_L$ decays immediately, in this scenario there is no $p\,p \to Q_L^*\,Q_L$ background with which to compete; also note, the Pati-Salam gauge boson-gluon coupling is through a suppressed dimension six operator.  Thus, there is no competition from $p\,p \to \psi^*\psi$ as well.  We still use the $\cos \phi$ cut to reduce the $Z \to \nu^* \nu$ background.  Applying all of the cuts listed in the previous section, we find signal events for 10 fb$^{-1}$ of integrated luminosity of
\begin{align}
S_\mathrm{ATLAS} = 12 	&& S_\mathrm{CMS} =  38.
\end{align}
In Figure 4a, we plot the missing energy for the signal for both ATLAS (solid) and CMS (dotted).  In Table II,  the signal significance and signal to background ratio is calculated. 

\subsection{Light Hidden Dark Sectors}

For this model, we repeat the same analysis as in the first subsection but with a 1 GeV long-lived gauge boson.  Again, we have assumed the hidden gauge coupling to be electroweak.   
\begin{figure}[t]
{\label{fig:costheta}
	\includegraphics[width=7truecm,height=5.2truecm,clip=true]{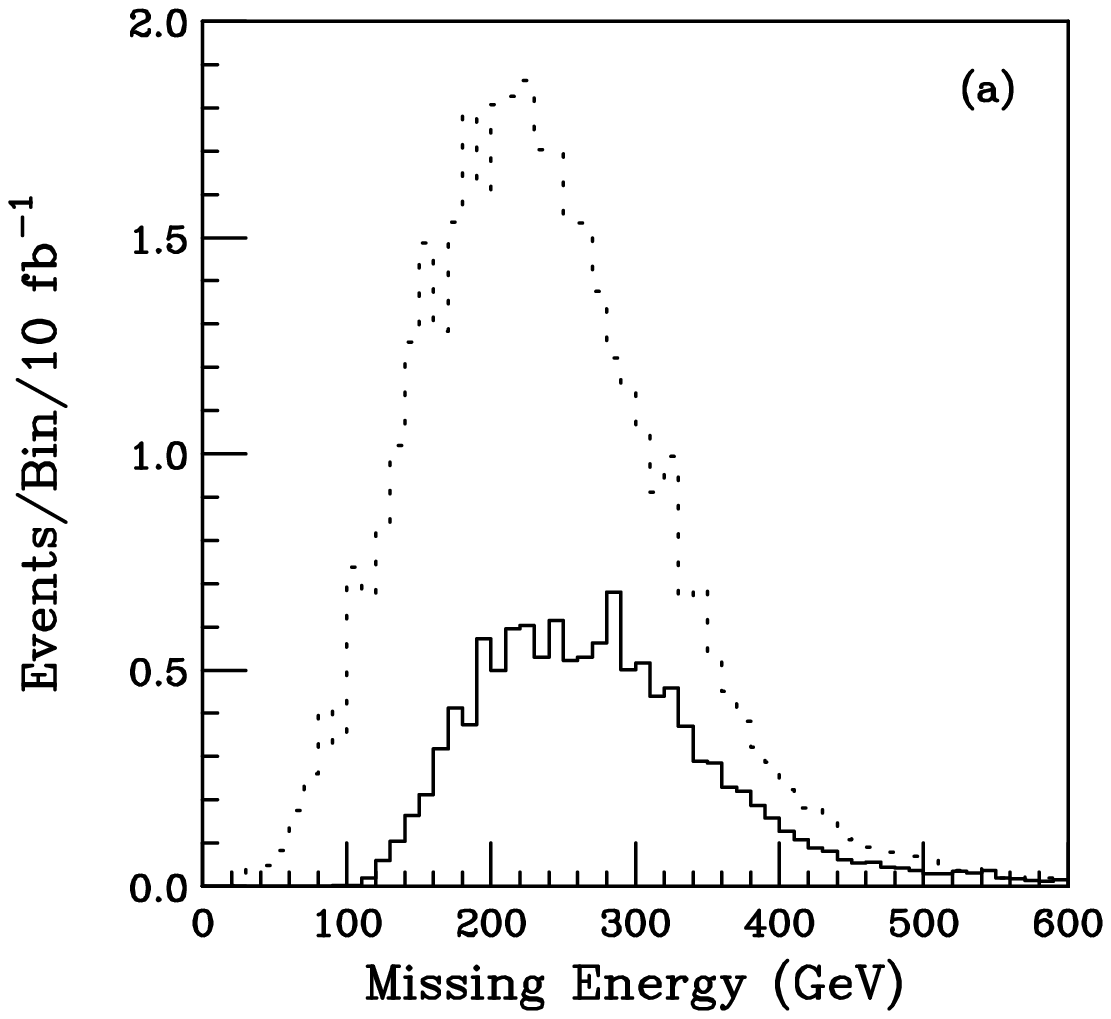}}
\vspace{0.3cm}
{\label{fig:costheta}
	\includegraphics[width=7truecm,height=5.2truecm,clip=true]{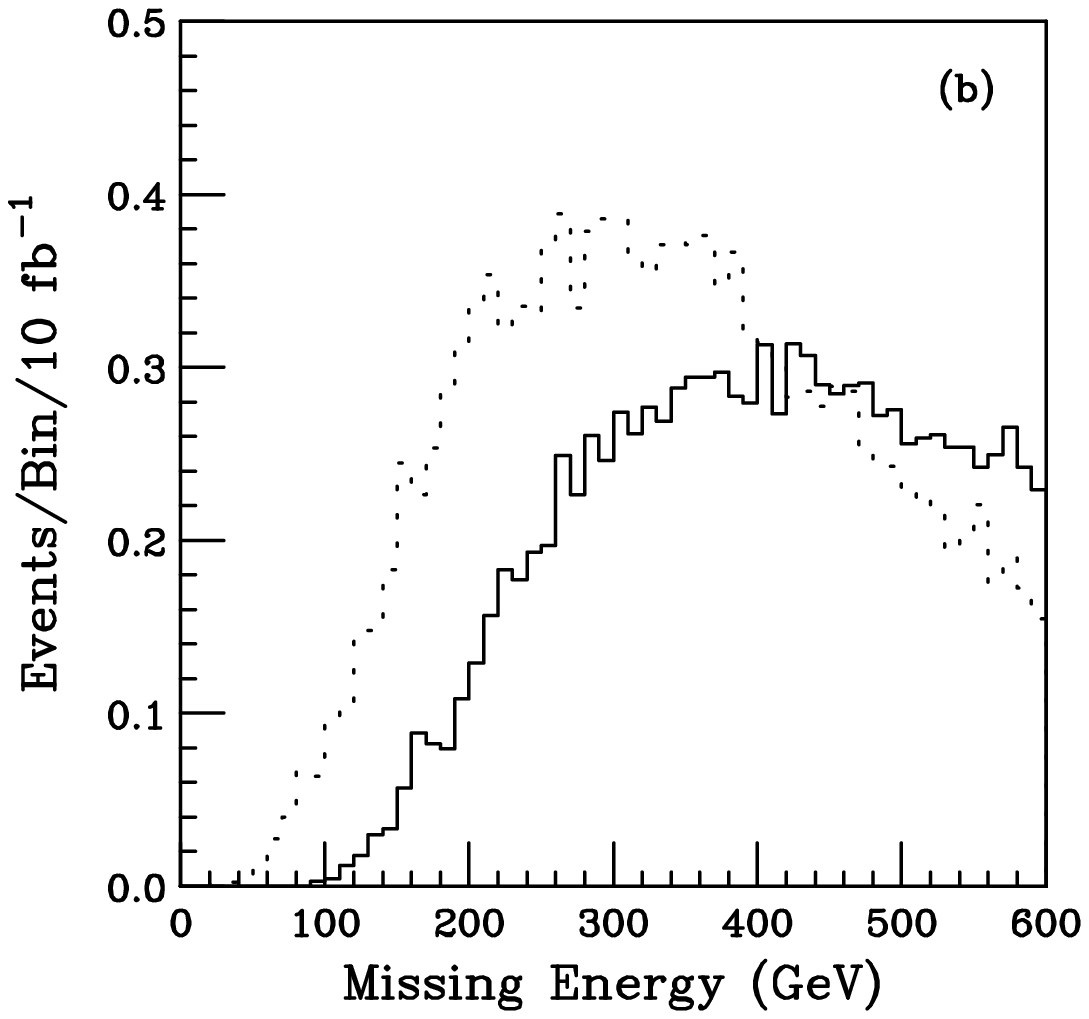}}	
\caption{Signal events vs. \!$\etmiss$  for ATLAS (solid) and CMS (dotted) in the AS scenario described in Section IIB (a) and the light hidden dark scenario described in Section IIC. (b)  In both plots, the $\cos \phi < -0.9$ cut is taken in addition to all of the kinematic cuts in Section III.}
\end{figure}
Because the dark gauge boson is much lighter, it does not force the $\psi$ particle go off-shell as much as in the companion model.  This increases the overall cross-section.  The larger cross section is down by a factor of $\alpha = g'^2/4\pi$ and is not competitive with the $p\,p \to \psi^* \psi$ background.  Although a 1 GeV light gauge boson does not  generally ``kick" the long-lived particles as much as the heavier dark gauge boson from the model in Section IIA, we still found it better to employ the $\cos \phi$ cut in order to reduce some of the SM background.  For 300 GeV $\psi$  and 10 fb$^{-1}$ the generated signal events are
\begin{align}
S_\mathrm{ATLAS} = 45	&& S_\mathrm{CMS} =  74.
\end{align}
For 600 GeV $\psi$
\begin{align}
S_\mathrm{ATLAS} = 1.85	&& S_\mathrm{CMS} =  2.52.
\end{align}
In Figure 4b, we plot the resulting missing energy.  The signal significance and the signal-to-background ratio is summarized for ATLAS and CMS in Table II.  A statistically significant signal should be possible for 100 fb$^{-1}$ of integrated luminosity for the case when $m_\psi = 300$ GeV.

\section{Summary}

In this letter we argued the signal of two long-lived particles plus large missing transverse energy were suppressed for models in which the dark matter and the long-lived particles were stabilized by a common parity ($\mathcal{Z}_2$) symmetry.  Models stabilized by a ``non-parity" symmetry may not be suppressed for this signal.  We used this process to distinguish between the two scenarios.  Specifically, we analyzed this signal process for three models in the literature which are stabilized by symmetries other than a parity.  For each we showed statistically significant signals for our choices in parameter space.  By comparison, an example parity symmetric effective theory generally contributed negligibly.  Finally, to generate reasonable estimates of the amount of missing energy generated by the missing dark matter candidates, we wrote a fast detector simulation to account for the energy loss effect for heavy long-lived particles interacting in the ATLAS and CMS detectors.
\vskip 0.3cm
\noindent

{\it Acknowledgments:}
We thank K.~Agashe, H.~Georgi, H.-S.~Goh, L.~Hall, I.~Hinchcliffe, A.E.~Nelson, V.~Sanz, M.~Shapiro, T.~Tait, S.~Vashen and P.~Zalewski for many useful discussions.  This work was supported by a University of California Presidential fellowship.  The author also acknowledges the hospitality of the Radcliffe Institute of Advanced Study at Harvard University during the earliest stages of this work.

\end{document}